\documentclass[conference]{IEEEtran}

\usepackage[pdftex]{graphicx}
\usepackage{url}
\usepackage[font=scriptsize]{caption}
\usepackage{multirow}

\begin{document}
	
	\renewcommand{\thetable}{\arabic{table}}
	\renewcommand{\figurename}{Fig.}

\title{Darknet and Deepnet Mining for Proactive Cybersecurity Threat Intelligence}

\author{\IEEEauthorblockN{Eric Nunes, Ahmad Diab,  Andrew Gunn, Ericsson Marin , Vineet Mishra,\\ Vivin Paliath, John Robertson,  Jana Shakarian, Amanda Thart, Paulo Shakarian}
	\IEEEauthorblockA{
		Arizona State University\\
		Tempe, AZ 85281, USA\\
		Email: \{enunes1, ahmad.diab, andrewgunn, ericsson.marin,  vvmishra,\\ vivin.paliath, jj.robertson, jshak, amanda.thart, shak\} @asu.edu}
	
	}

\maketitle

\begin{abstract}
In this paper, we present an operational system for cyber threat intelligence gathering from various social platforms on the Internet particularly sites on the darknet and deepnet. We focus our attention to collecting information from hacker forum discussions and marketplaces offering products and services focusing on malicious hacking. We have developed an operational system for obtaining information from these sites for the purposes of identifying emerging cyber threats.  Currently, this system collects on average 305 high-quality cyber threat warnings each week.  These threat warnings include information on newly developed malware and exploits that have not yet been deployed in a cyber-attack. This provides a significant service to cyber-defenders. The system is significantly augmented through the use of various data mining and machine learning techniques. With the use of machine learning models, we are able to recall 92\% of products in marketplaces and 80\% of discussions on forums relating to malicious hacking with high precision. We perform preliminary analysis on the data collected, demonstrating its application to aid a security expert for better threat analysis.
\end{abstract}

\IEEEpeerreviewmaketitle

\section{Introduction}
\label{intro}
Pre-reconnaissance cyber threat intelligence refers to information gathered before a malicious party interacts with the defended computer system.  An example demonstrating the importance of cyber threat intelligence is shown in Table~\ref{exploit}. A Microsoft Windows vulnerability was identified in Feb. 2015. The release of the vulnerability was essentially Microsoft warning its customers of a security flaw.  Note that at this time, there was no publicly known method to leverage this flaw  in a cyber-attack (i.e. an available exploit). However, about a month later an exploit was  found to be on sale in darknet market. It was not until July when FireEye, a major cybersecurity firm, identified that the Dyre Banking Trojan designed to steal credit cards exploited this vulnerability - the first time an exploit was reported.  This vignette demonstrates how threat warnings gathered from the darknet can provide valuable information for security professionals. The average global exposure of the Dyre Banking Trojan was 57.3\% along with another banking malware Dridex\footnote{https://www.fireeye.com/blog/threat-research/2015/06/evolution\_of\_dridex.html}. It means that nearly 6 out of 10 organizations in the world were affected, and this is a significantly high number on a global level. 

In this paper, we examine how such intelligence can be gathered and analyzed from various social platforms on the Internet particularly sites on the darknet and deepnet. In doing so, we encounter several problems that we addressed with various data mining techniques.  Our current system is operational and actively collecting approximately $305$ cyber threats each week. 
Table~\ref{status} shows the current database statistics. It shows the total data collected and the data related to malicious hacking. The vendor and user statistics cited only consider those individuals associated in the discussion or sale of malicious hacking-related material, as identified by the system. The data is collected from two sources on the darknet/deepnet: markets and forums. 
\begin{table}[t!]
	\caption{\textmd{Exploit example.}}
	\scriptsize
	\label{exploit}
	\centering
	\renewcommand{\arraystretch}{1.5}
	
	\begin{tabular}{|p{1.6cm}|p{5.5cm}|}
		\hline
		{\bf Timeline} &  {\bf Event} \\ \hline
		
		Feb. 2015 & Microsoft identifies Windows vulnerability MS15-010/CVE 2015-0057 for remote code execution. There was no publicly known exploit at the time the vulnerability was released.  \\ \hline
		April 2015 & An exploit for MS15-010/CVE 2015-0057 was found on a darknet market on sale for 48 BTC (around \$10,000-15,000). \\ \hline
		July 2015 & FireEye identified that the Dyre Banking Trojan, designed to steal credit card number, actually exploited this vulnerability$^1$. \\ \hline

	\end{tabular}
	
\end{table}
\begin{table}[htb!]
	\caption{\textmd{Current Database Status}}
	\scriptsize
	\label{status}
	\centering
	\renewcommand{\arraystretch}{1.5}
	
	\begin{tabular}{ |l|l|l| }
		\hline
		\multirow{4}{*}{Markets} & Total Number & 27 \\
		& Total products & 11991  \\
		& Hacking related & 1573 \\
		& Vendors & 434 \\ \hline
		\multirow{3}{*}{Forums} & Total Number & 21 \\
		& Topics/Posts & 23780/162872 \\
		& Hacking related & 4423/31168 \\ 
		& Users & 5491 \\ \hline
	
	\end{tabular}
	
\end{table}

We are providing this information to cyber-security professionals to support their strategic cyder-defense planning to address questions such as, \textit{1) What vendors and users have a presence in multiple darknet/deepnet markets/ forums?} \textit{2)What zero-day exploits are being developed by malicious hackers?} \textit{3) What vulnerabilities do the latest exploits target?}\\

Specific contributions of this paper include, 1) Description of a system for cyber threat intelligence gathering from various social platforms from the Internet such as deepnet and darknet websites. 2) The implementation and evaluation of learning models to separate relevant information from noise in the data collected from these online platforms. 3) A series of case studies showcasing various findings relating to malicious hacker behavior resulting from the data collected by our operational system.

\noindent\textbf{Background:}
\label{back}
Many of the individuals behind cyber-operations -- originating outside of government run labs or military commands -- rely on a significant community of hackers. They interact through a variety of online forums (as means to both stay anonymous and to reach geographically dispersed collaborators). 

\noindent\textit{Darknet and Deepnet Sites:}
Widely used for underground communication, ``The Onion Router'' (Tor) is free software dedicated to protect the privacy of its users by obscuring traffic analysis as a form of network surveillance \cite{Dingledineetal2004}. The network traffic in Tor is guided through a number of volunteer-operated servers (also called ``nodes''). Each node of the network encrypts the information it blindly passes on neither registering where the traffic came from nor where it is headed \cite{Dingledineetal2004}, disallowing any tracking. Effectively, this allows not only for anonymized browsing (the IP-address revealed will only be that of the last node), but also for circumvention of censorship\footnote{See the Tor Project's official website (https://www.torproject.org/)}. Here, we will use ``darknet'' to denote the anonymous communication provided by crypto-networks like ``Tor'', which stands in contrast to ``deepnet'' which commonly refers to websites hosted on the open portion of the Internet (the ``Clearnet''), but not indexed by search engines \cite{LaceySalmon2015}. 

\noindent\textit{Markets:} Users advertise and sell their wares on marketplaces. Darknet marketplaces provide a new avenue to gather information about the cyber threat landscape. The marketplaces sell goods and services relating to malicious hacking, drugs, pornography, weapons and software services. Only a small fraction of products (13\% in our collected data to date) are related to malicious hacking. Vendors often advertise their products on forums to attract attention towards their goods and services.  

\noindent\textit{Forums.} Forums are user-oriented platforms that have the sole purpose of enabling communication. It provides the opportunity for the emergence of a community of like-minded individuals - regardless of their geophysical location. Administrators set up Darknet forums with communication safety for their members in mind. While structure and organization of Darknet-hosted forums might be very similar to more familiar web-forums, the topics and concerns of the users vary distinctly. Forums addressing malicious hackers feature discussions on programming, hacking, and cyber-security. Threads are dedicated to security concerns like privacy and online-safety - topics which plug back into and determine the structures and usage of the platforms.

\section{SYSTEM OVERVIEW}
\label{sys}
\begin{figure*}[!t]
	\centerline{\includegraphics[width=0.8\textwidth,keepaspectratio]{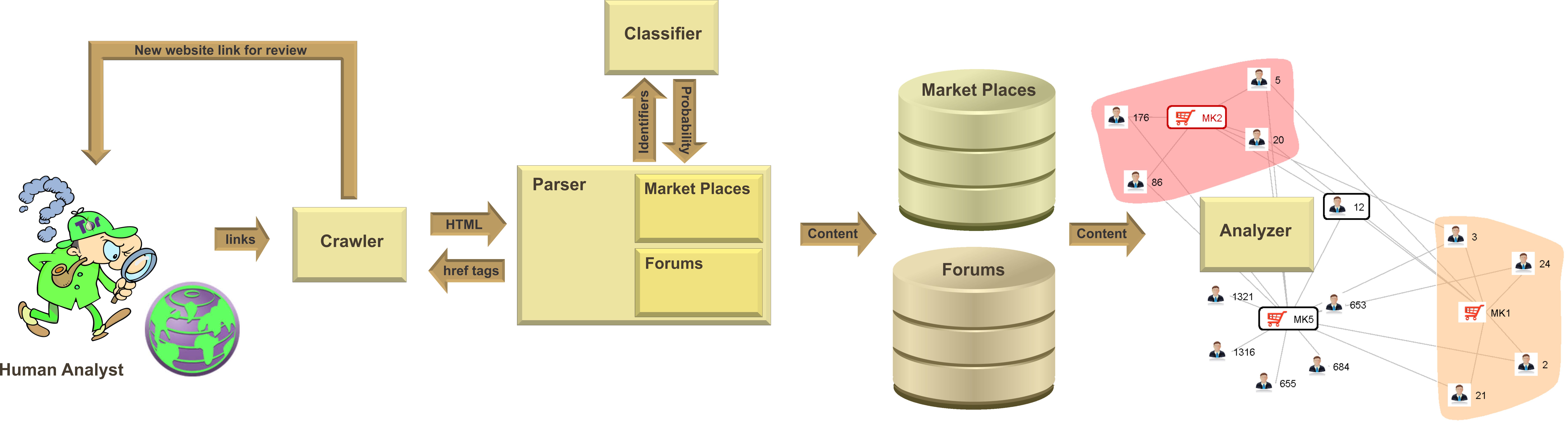}}
	\caption{\textmd{System overview}}
	\label{system}
\end{figure*}

\figurename~\ref{system} gives the overview of the system. Through search engines and spider services on the Tor network, human analysts were able to find forums and marketplaces populated by malicious hackers. Other platforms were discovered through links posted on forums either on the Tor-network or on the Clearnet. The system consists of three main modules built independently before integration. The system is currently fully integrated and actively collecting cyber threat intelligence.\\
\textbf{Crawler:} The crawler is a program designed to traverse the website and retrieve HTML documents. Topic based crawlers have been used for focused crawling where only webpages of interest are retrieved~\cite{menczer2004topical,chakrabarti1999focused}. More recently, focused crawling was employed to collect forum discussions from darknet~\cite{fu2010focused}. We have designed separate crawlers for different platforms (markets/forums) identified by experts due to the structural difference and access control measures for each platform. In our crawler, we address design challenges like accessibility, unresponsive server, repeating links creating a loop etc. to gather information regarding products from markets and discussions on forums.\\
\textbf{Parser:} We designed a parser to extract specific information from marketplaces (regarding sale of malware/exploits) and hacker forums (discussion regarding services and threats). This well-structured information is stored in a relational database. We maintain two databases, one for marketplaces and the other for forums. Like the crawler, each platform has its own parser. The parser also communicates with the crawler from time to time for collection of temporal data. The parser communicates a list of relevant webpages to the crawler, which are re-crawled to get time-varying data. For markets we collect the following important products fields: \textit{\{item\_title, item\_description, vendor\_name, shipping\_details, item\_reviews, items\_sold, CVE, items\_left, 
	transaction\_details, ratings\}}. For forums we collect the following fields: \textit{\{topic\_content, post\_content, topic\_author, post\_author, author\_status, reputation, topic\_interest\}}.\\
\textbf{Classifier:}
We employ a machine learning technique using an expert-labeled dataset to detect relevant products and topics from marketplaces and forums respectively discussed in Section~\ref{eval}. These classifiers are integrated into the parser to filter out products and topics relating to drugs, weapons, etc. not relevant to malicious hacking.

\section{Evaluation}
\label{eval}
We consider the  classification of identifying relevant products in darknet/deepnet marketplaces and relevant topics on  forum post containing  communication relevant to malicious hacking in this paper. It is a binary classification problem with the data sample (in this case products/forum topics) being relevant or not. We look at both supervised and semi-supervised approaches to address the classification.

\subsection{Machine Learning Approaches}
In this work, we leverage a combination of supervised and semi-supervised methods.  Supervised methods include the well-known classification techniques of Naive Bayes (NB), random forest (RF), support vector machine (SVM) and logistic regression (LOG-REG).  However, supervised techniques required labeled data, and this is expensive and often requires expert knowledge. Semi-supervised approaches work with limited labeled data by leveraging information from unlabeled data. We discuss popular semi-supervised approaches used in this work. We perform a grid search to find optimal parameters for the learning techniques.\\

\noindent\textbf{Label propagation (LP).} The label propagation approach \cite{Zhu03combiningactive} has been widely used for semi-supervised classification task \cite{BishopU04,Levin:2006,WangYZZ09,ChengLY09}. It estimates the label values based on  graph Laplacian \cite{nips2002:belkin} where the model is represented by a weighted graph $ G = (V , E)$ , where $V$ indicates the vertices representing the samples, while the edges E are the weights indicating the similarity between points. A subset of these vertices are labeled and these vertices are then used to estimate the labels of the remaining under the assumption that the edges are able to capture the similarity between samples. Hence, the performance of these methods depends on the similarity measure used. The most commonly used similarity measures include $k$-NN and Gaussian kernel.

\noindent\textbf{Co-training (CT).}  The Co-training approach was proposed by Blum and Mitchell \cite{Blum:1998}. In this approach, the feature set is divided into two sets (assumed to be independent), and two classifiers are trained using the limited labeled set denoted by $L$ . These trained classifiers are then used to estimate the labels for the unlabeled points. High confidence label estimates from classifier-1 are added to the labeled set $L$ of classifier-2 and vice versa. For the current setting we set the confidence to 70\%. Every time the labeled set $L$ is updated, the classifiers are retrained. This procedure repeats until all of the unlabeled points are labeled. It can be viewed as two classifiers teaching each other. 

\subsection{Experiments: Marketplaces}
Marketplaces sell goods and services that do not relate to malicious hacking, including drugs, pornography, weapons and software services. Only a small fraction of products (13\%) are related to malicious hacking. We thus require a model that can separate relevant products from the non-relevant ones. The data collected from marketplaces is noisy and hence not suitable to use directly as input to a learning model. Hence, the raw information undergoes several steps of automated data cleaning. We now discuss the challenges associated with the dataset obtained and the data processing steps taken to address them. We note that similar challenges occur for forum data.

\noindent\textbf{Text Cleaning.}
Product title and descriptions on marketplaces often have much text that serves as noise to the classifier (e.g. *****SALE*****).  To deal with these instances, we first removed all non-alphanumeric characters from the title and description.  This, in tandem with standard stop-word removal, greatly improved classification performance.

\noindent\textbf{Misspellings and Word Variations.} 
Misspellings frequently occur on forums and marketplaces, which is an obstacle for the standard bag-of-words classification approach.  Additionally, with the standard bag-of-words approach, variations of words are considered separately (e.g. hacker, hack, hackers, etc.).  Word stemming mitigates these issue of word variations, but fails to fix the issue of misspellings.  To address this we use character $n$-gram features.  As an example of character $n$-gram features, consider the word ``hacker".  If we were using tri-gram character features, the word ``hacker" would yield the features ``hac'', ``ack'', ``cke'', ``ker''.  The benefit of this being that the variations or misspellings of the word in the forms ``hack", ``hackz", ``"hackker", will all have some common features.  We found that using character $n$-grams in the range (3, 7) outperformed word stemming in our experiments.

\noindent\textbf{Large Feature Space.} 
In standard bag-of-words approach, as opposed to the character $n$-gram approach, the feature matrix gets very large as the number of words increase. As the number of unique words grow, this bloated feature matrix begins to greatly degrade performance.  Using $n$-gram features further increases the already over-sized feature matrix.  To address this issue, we leveraged the sparse matrix data structure in the scipy\footnote{http://www.scipy.org/} library, which leverages the fact that most of the entries will be zero.  If a word or $n$-gram feature is not present in a given sample, there is simply no entry for that feature in the sparse matrix.  

\noindent\textbf{Preserving Title Feature Context.} 
As the title and description of the product are disjoint, we found that simply concatenating the description to the product title before extracting features led to sub-optimal classification performance. We believe that by doing a simple concatenation, we were losing important contextual information. There are features that should be interpreted differently should they appear in the title versus the description.  Initially, we used two separate classifiers: one for the title and one for the description.  With this construction, when an unknown product was being classified, we would pass the title to the title classifier and the description to the description classifier.  If either classifier returned a positive classification, we would assign the product a positive classification. However, we believe that this again led to the loss of important contextual information. To fix this, we independently extract character $n$-gram features from the title and description. This step yields a title feature vector and a description feature vector. We then horizontally concatenate these vectors, forming a single feature vector which includes separate feature sets for the title and description.

\begin{figure}[htp!]
	\centerline{\includegraphics[scale=0.2,keepaspectratio]{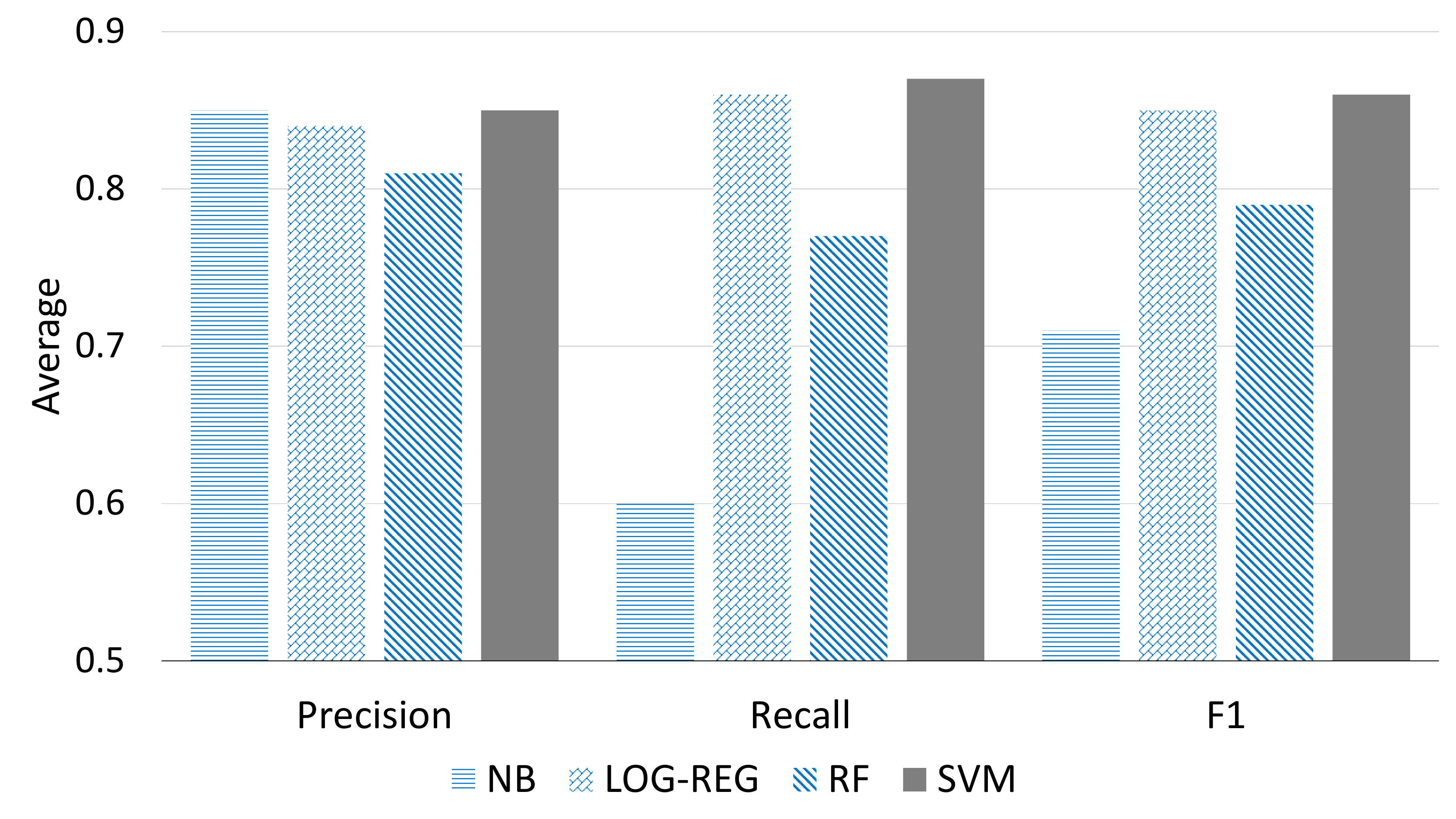}}
	\caption{\textmd{Average Precision, Recall and F1 comparisons for NB, LOG-REG, RF and SVM for product classification.}}
	\label{fig:one}
\end{figure}

\noindent\textbf{Results:}
We consider 10 marketplaces to train and test our learning model. A summary of these marketplaces is shown in Table~\ref{market}. Table~\ref{example} gives an instance of products defined as being relevant or not. With the help of security experts we label 25\% of the products from each marketplace. The experimental setup is as follows. We perform a leave-one-marketplace-out cross-validation.  In other words, given $n$ marketplaces we train on $n-1$ and test on the remaining  one. We repeat this experiment for all the marketplaces. For the supervised experiment, we only use the 25\% labeled data from each marketplace. We evaluate the performance based primarily on three metrics: precision, recall and unbiased F1. Precision indicates the fraction of products that were relevant from the predicted ones. Recall is the fraction of relevant products retrieved. F1 is the harmonic mean of precision and recall. The results are averaged and weighted by the number of samples in each market. In this application, a high recall is desirable as we do not want to omit relevant products. In the supervised approaches, SVM with linear kernel performed the best, recalling 87\% of the relevant products while maintaining a precision of 85\% (\figurename~\ref{fig:one}). SVM performed the best likely due to the fact it maximizes generality as opposed to minimizing error.

\begin{table}[t]
	\caption{\textmd{Markets and Number of products collected.}}
	\scriptsize
	\label{market}
	\begin{minipage}{.4\linewidth}
		\centering
		\renewcommand{\arraystretch}{1.5}
		
		\begin{tabular}{|l|l|}
			\hline
			{\bf Markets} &  {\bf Products} \\ \hline
			
			Market-1 & 439 \\ \hline
			Market-2 & 1329 \\ \hline
			Market-3 & 455 \\ \hline
			Market-4 & 4018 \\ \hline
			Market-5 & 876 \\ \hline
			
		\end{tabular}
	\end{minipage}\qquad \quad
	\begin{minipage}{.4\linewidth}
		\centering
		\renewcommand{\arraystretch}{1.5}
		\begin{tabular}{|l|l| }
			\hline
			{\bf Markets} &  {\bf Products} \\ \hline
			Market-6 & 497 \\ \hline
			Market-7 & 491 \\ \hline
			Market-8 & 764 \\ \hline
			Market-9 & 2014 \\ \hline
			Market-10 & 600 \\ \hline

		\end{tabular}
	\end{minipage}
\end{table}

\begin{table}[htb!]
	\caption{\textmd{Example of Products.}}
	\scriptsize
	\label{example}
	\centering
	\renewcommand{\arraystretch}{1.5}
	
	\begin{tabular}{|p{6cm}|p{1.4cm}|}
		\hline
		{\bf Product Title} &  {\bf Relevant} \\ \hline
		
		20+ Hacking Tools (Botnets  Keyloggers  Worms and More!) & YES \\ \hline
		5 gm Colombian Cocaine &  NO \\ \hline
		
	\end{tabular}

\end{table}

As stated, only 25\% of the data is labeled, as labeling often requires expert knowledge. However, this significant cost and time investment can be reduced by applying a semi-supervised approach which leverages the unlabeled data to aid in classification. It takes approximately one minute for a human to label 5 marketplace products or 2 topics on forums as relevant or not, highlighting the costliness of manual labeling. The experimental setup is similar to the supervised approach, but this time we also utilize the large unlabeled data from each marketplace (75\%) for training. 

\begin{figure}[htp!]
	\centerline{\includegraphics[scale=0.2,keepaspectratio]{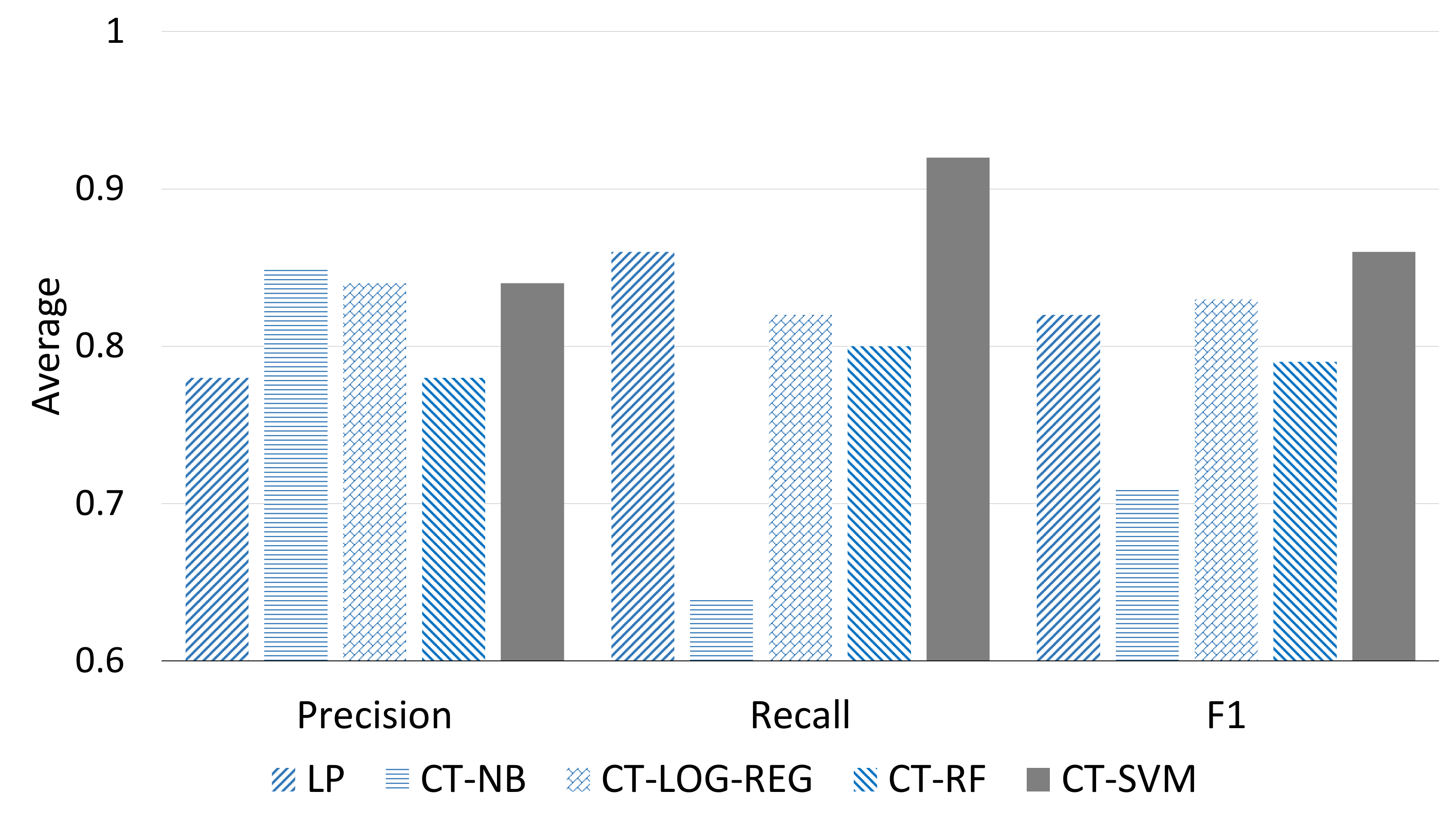}}
	\caption{\textmd{Average Precision, Recall and F1 comparisons for LP, CT-NB, CT-LOG-REG, CT-RF and CT-SVM for product classification.}}
	\label{fig:two}
\end{figure}
\figurename~\ref{fig:two} shows the performance comparison for the semi-supervised approaches. For the co-training approach, we divide the feature space into two sets. The two feature sets used  are both based on character n-grams.  However, the set of words from which the character n-grams are derived are disjoint between the two sets.  In this way, the two corresponding feature vectors  can be treated as being independent from one another. Hence we get two views of the same sample. Co-training with Linear SVM is able to recall 92\% of the relevant products as compared to label propagation and other variants of co-training while maintaining a precision of 82\%, which is desirable. In this case, the unlabeled data aided the classification in improving the recall to 92\% without significantly reducing the precision.

\subsection{Experiment: Forums}
In addition to the darknet/deepnet marketplaces that we have already discussed, there are also numerous darknet forums on which users discuss malicious hacking related topics. Again, there is the issue that only a fraction of these topics with posts on these forums contain information that is relevant to malicious hacking or the trading of exploits. Hence, we need a classifier to identify relevant topics. This classification problem is very similar to the product classification problem previously discussed, with similar set of challenges. 

\begin{figure}[htp!]
	\centerline{\includegraphics[scale=0.2,keepaspectratio]{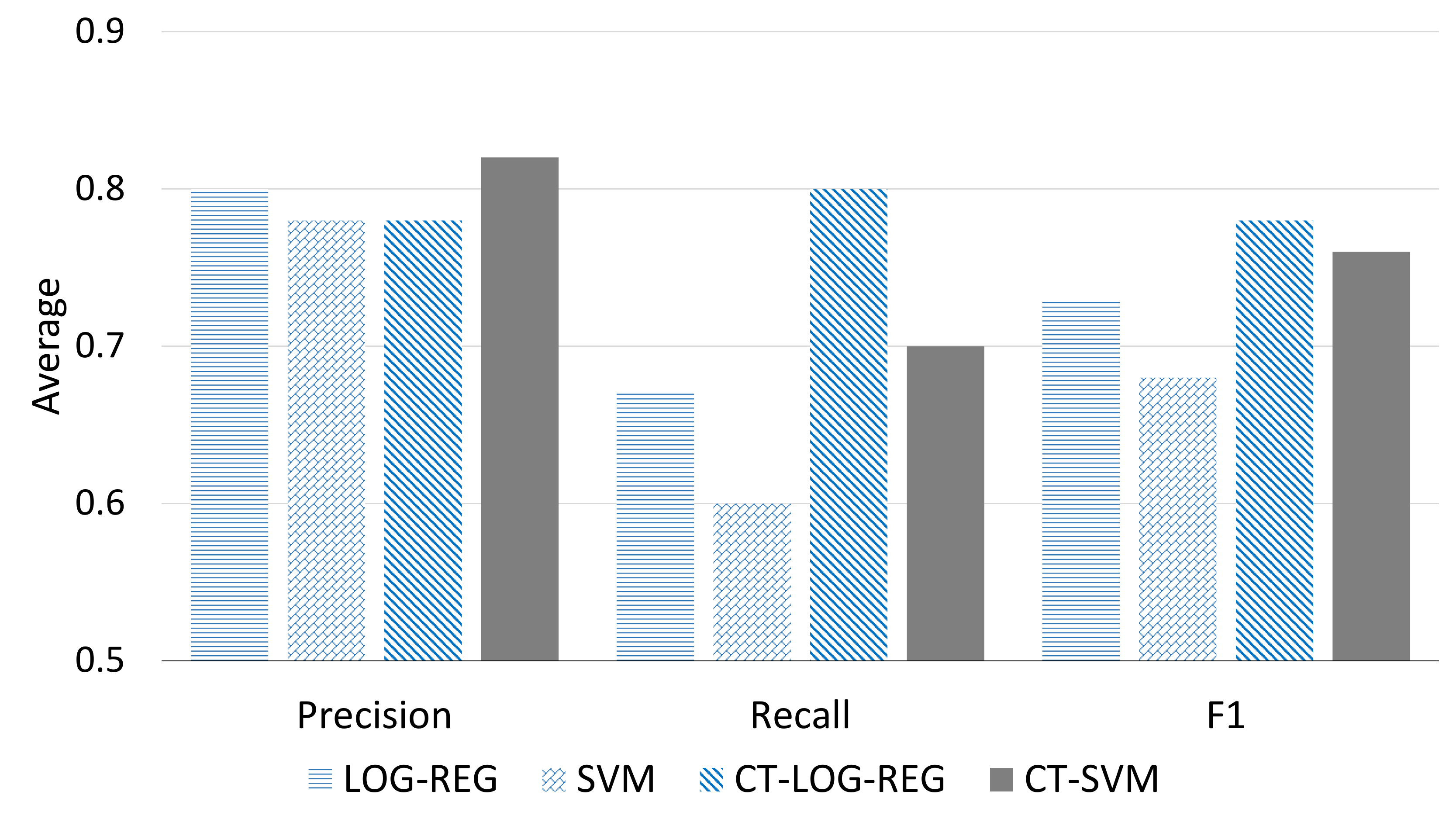}}
	\caption{\textmd{Average Precision, Recall and F1 comparisons for LOG-REG, SVM, CT-LOG-REG, and CT-SVM for English forum topic classification.}}
	\label{Eforum}
\end{figure}

We performed evaluation on two such English forums. The dataset consisted of 781 topics with 5373 posts. Table~\ref{exampleF} gives instance of topics defined as being relevant or not. We label 25\% of the topics and perform a 10-fold cross validation using supervised methods. We show the results from the top two performing supervised and semi-supervised methods. In the supervised setting, LOG-REG performed the best with 80\% precision and 68\% recall (\figurename~\ref{Eforum}). On the other hand, leveraging unlabeled data in a semi-supervised technique improved the recall while maintaining the precision. We note that in this case the 10-fold cross validation was performed only on the labeled points. In the semi-supervised domain co-training with LOG-REG improved the recall to 80\% with precision of 78\%.

\begin{table}[htb!]
	\caption{\textmd{Example of Topics.}}
	\scriptsize
	\label{exampleF}
	\centering
	\renewcommand{\arraystretch}{1.5}
	
	\begin{tabular}{|p{5cm}|p{2cm}|}
		\hline
		{\bf Topic} &  {\bf Relevant} \\ \hline
		Bitcoin Mixing services & YES \\ \hline
		Looking for MDE/MDEA shipped to Aus &  NO \\ \hline
		
	\end{tabular}

\end{table} 

\section{Case Studies}
\label{case}
We analyze the data with the purpose of answering the questions raised in the Section~\ref{intro}. We will be using the following key security terms. \textit{Vulnerability} is a security flaw that allows an attacker to compromise a software or an operating system. \textit{Exploit} is a piece of software that takes advantage of a vulnerability in a piece of software or operating system to compromise it. \textit{Patch} is a piece of software used to improve existing software by fixing vulnerabilities to improve security. We discuss the following case-studies.

\subsection{Discovery of Zero-Day Exploits.}
Over a 4 week period, we detected 16 zero-day exploits from the marketplace data. Zero-day exploits  leverage vulnerabilities that are unknown to the vendor. Table~\ref{0day} shows a sample of zero-day exploits with their selling price in Bitcoin. The Android WebView zero-day affects a vulnerability in the rendering of web pages in Android devices. It affects devices running on Android 4.3 Jelly Bean or earlier versions of the operating system. This comprised of more than 60\% of the Android devices in 2015. After the original posting of this zero-day, a patch was released in Android KitKit 4.4 and Lollipop 5.0 which required devices to upgrade their operating system. As not all users have/will update to the new operating system, the exploit continues to be sold for a high price. Detection of these zero-day exploits at an earlier stage can help organizations avoid an attack on their system or minimize the damage. For instance, in this case, an organization may decide to prioritize patching, updating, or replacing certain systems using the Android operating system. 
\begin{table}[htb!]
	\caption{\textmd{Example of Zero-day exploits.}}
	\scriptsize
	\label{0day}
	\centering
	\renewcommand{\arraystretch}{1.5}
	
	\begin{tabular}{|p{5cm}|p{2.3cm}|}
		\hline
		{\bf Zero-day exploit} &  {\bf Price (BTC)} \\ \hline
		
		Internet Explorer 11 Remote Code Execution 0day  & 20.4676 \\ \hline
		Android WebView 0day RCE &  40.8956 \\ \hline

	\end{tabular}

\end{table}

\subsection{Users having presence in markets/ forums.}
\begin{figure}[htp!]
	\centerline{\includegraphics[scale=0.15,keepaspectratio]{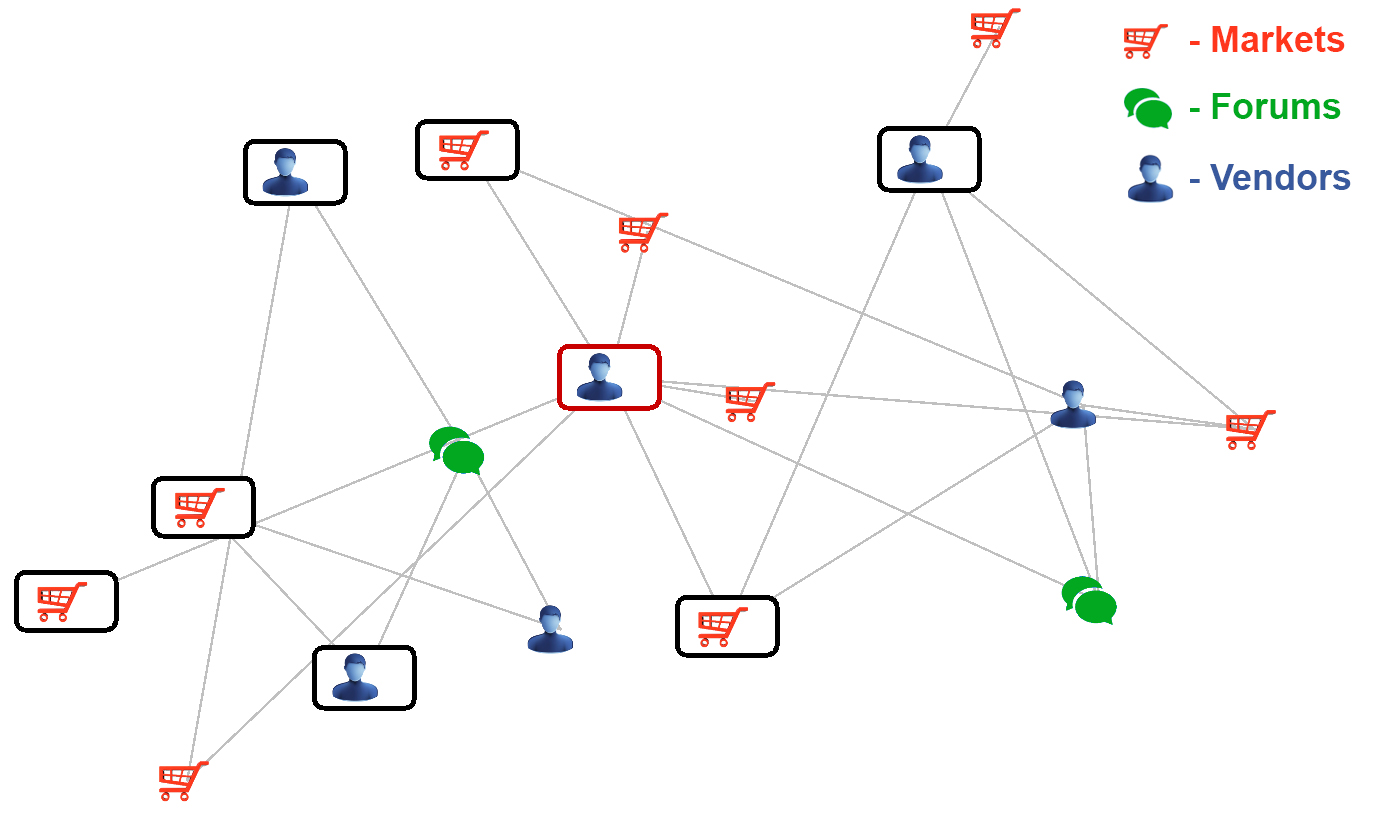}}
	\vspace{-0.5em}
	\caption{\textmd{Vendor/User network in marketplace and forum.}}
	
	\label{net1}
\end{figure}

\begin{figure}[htp!]
	\centering
	\begin{minipage}{.45\linewidth}
		\includegraphics[width=\linewidth]{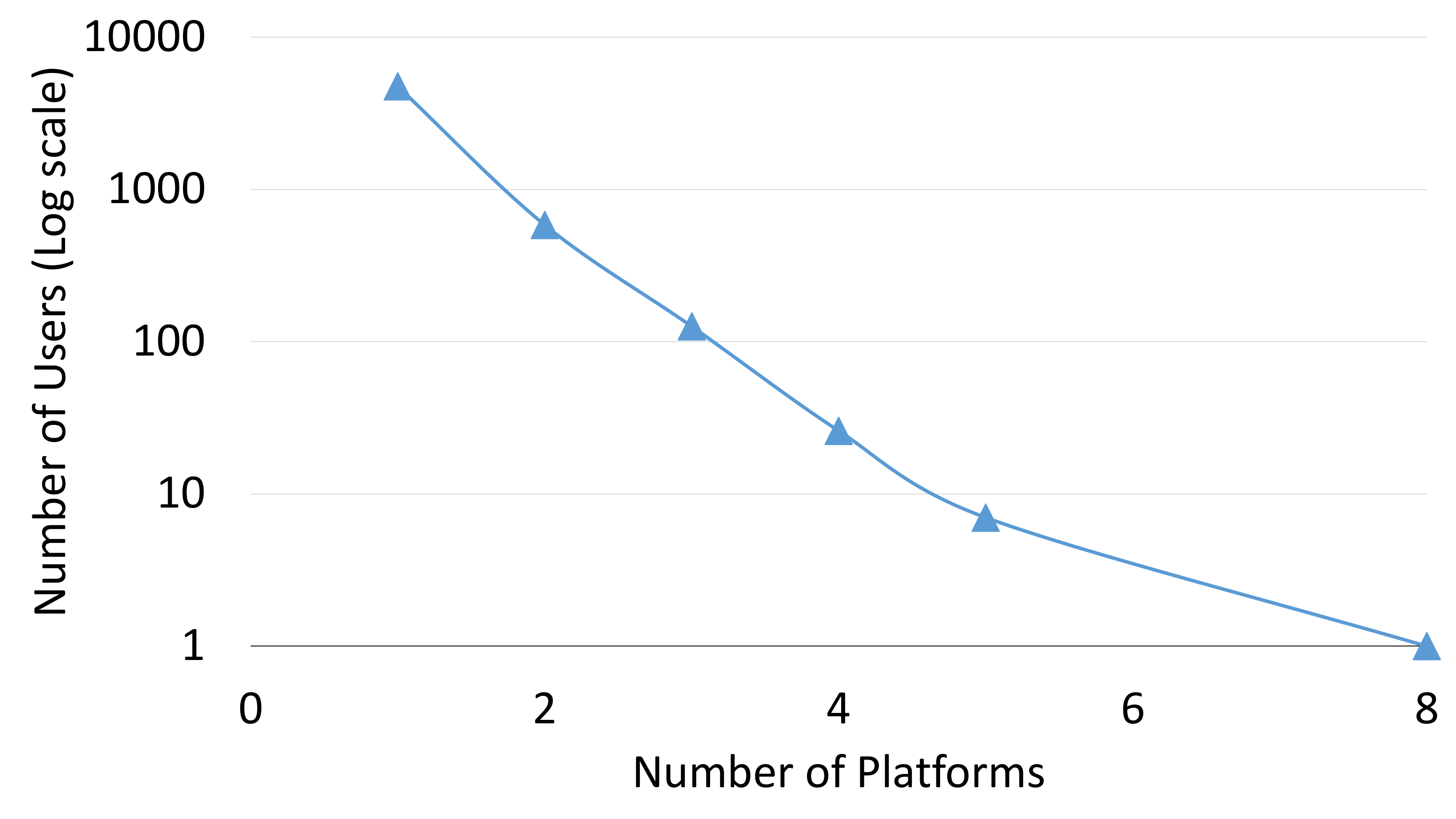}
		\caption{\textmd{Users in multiple markets and forums.}}
		\label{users}
	\end{minipage}
	\hspace{.05\linewidth}
	\begin{minipage}{.45\linewidth}
		\includegraphics[width=\linewidth]{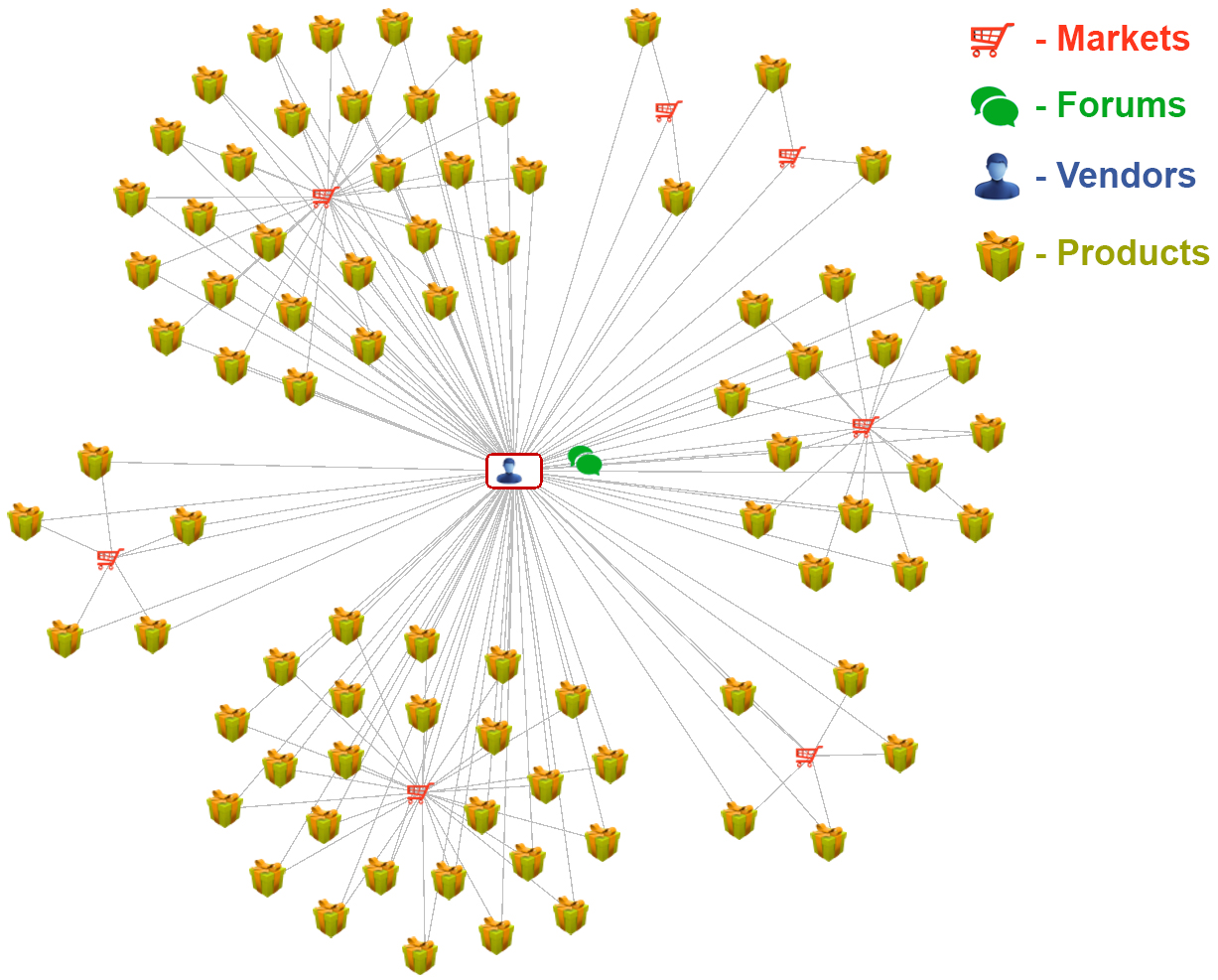}
		\caption{\textmd{A centric network of a Vendor.}}
		\label{net2}
	\end{minipage}
\end{figure}

Previous studies on darknet crawling~\cite{fu2010focused,benjamin2015exploring} explore a single domain, namely forums. We create a social network that includes both types of information studied in this paper: marketplaces and forums. We can thus study and find these cross-site connections that were previously unstudied. We are able to produce this connected graph using the ``usernames'' used by vendors and users in each domain. A subgraph of this network containing some of the individuals who are simultaneously selling products related to malicious hacking and publishing in hacking related forums is shown in \figurename~\ref{net1}. In most cases, the vendors are trying to advertise/discuss their products on the forums, demonstrating their expertise. Using these integrated graphic representations, one can visualize the individuals' participation in both domains, making the right associations that lead to a better comprehension of the malicious hacker networks. It is helpful in determining social groups within the forums of user interaction. The presence of users on multiple markets and forums follows a power law. From \figurename~\ref{users}, majority of users only belong to a single market or forum. We note that there are 751 users that are present in more than two platforms. \figurename~\ref{net2} considers one such user/vendor. The vendor is active in 7 marketplaces and 1 forum . The vendor offers 82 malicious hacking related products and discusses these products on the forum. The vendor has an average rating of 4.7/5.0, rated by customers on the marketplace with more than 7000 successful transactions, indicating the reliability of the products and the popularity of the vendor. 

\section{Related Work}
\label{related}
Web crawling is a popular way of collecting large amounts of data from the Internet. In many applications, researchers are interested in specific topics for their application. Hence, the need for a topic-based crawler popularly referred to as a focused crawler~\cite{chakrabarti1999focused,chakrabarti2002accelerated}. Most of the focused crawlers are designed to collect information from the \textit{surface web} with little concentration on the darknet websites. More recently, a focused crawler concentrating on dark web forums was designed~\cite{fu2010focused}. This research primarily concentrated on forums, collecting data over a period of time and then performing static analysis to study online communities. The authors also describe different data mining techniques for these forums in~\cite{chen2011dark}. We, on the other hand, not only look at darknet forums but also collect information from marketplaces hosting a range of products relating to malicious hacking. 
Another application of leveraging darknet information to counter human trafficking is developed by DARPA through the Memex program\footnote{http://opencatalog.darpa.mil/MEMEX.html} - a program with different goals than the work described in this paper. 

Previous work leverages the exploit information from marketplaces in a game theoretic framework to formulate system configurations that minimize the potential damage of a malicious cyber attack~\cite{JJgame16}. Work analyzing hacker forums to detect threats that pose great risk to individuals, businesses, and government have been discussed in~\cite{benjamin2015exploring}. It further states that knowledge is distributed in forums. That minimally skilled people could learn enough by simply frequenting such platforms. Studying these hacker communities gives insights in the social relationships. Also, the distribution of information amongst users in these communities based on their skill level and reputation~\cite{holt2012examining,jordan1998sociology,holt2007subcultural}. These forums also serve as markets where malware and stolen personal information are shared / sold \cite{holt2010exploring}. Samtani et al. analyze hacker assets in underground forums~\cite{samtani2015exploring}. They discuss the dynamics and nature of sharing of tutorials, source code, and ``attachments'' (e.g. e-books, system security tools, hardware/software). Tutorials appear to be the most common way of sharing resources for malicious attacks. Source code found on these particular forums was not related to specific attacks. Additionally underground (not malicious hacking related) forums have also been analyzed to captures the dynamic trust relationships forged between mutually distrustful parties~\cite{motoyama2011analysis}.

\section{Conclusion}
\label{con}
In this paper, we implement a system for intelligence gathering related to malicious hacking. Our system is currently operational.  We are in the process of transitioning this system to a commercial partner. We consider social platforms on darknet and deepnet for data collection.  We address various design challenges to develop a focused crawler using data mining and machine learning techniques. The constructed database is made available to security professionals in order to identify emerging cyber-threats and capabilities.

\begin{small}
\noindent\textbf{Acknowledgments:} Some of this work is supported by ONR NEPTUNE, ASU GSI, ASU ISSR and CNPq-Brazil.
\end{small}

\bibliographystyle{abbrv}

\bibliography{ref}

\end{document}